\documentstyle[citesort,epsfig,twoside]{article}

\catcode`\@=11
\long\def\@makefntext#1{
\protect\noindent \hbox to 3.2pt {\hskip-.9pt  
$^{{\eightrm\@thefnmark}}$\hfil}#1\hfill}               

\def\@makefnmark{\hbox to 0pt{$^{\@thefnmark}$\hss}}    
        
\def\ps@myheadings{\let\@mkboth\@gobbletwo
\def\@oddhead{\hbox{}
\rightmark\hfil\eightrm\thepage}   
\def\@oddfoot{}\def\@evenhead{\eightrm\thepage\hfil
\leftmark\hbox{}}\def\@evenfoot{}
\def\sectionmark##1{}\def\subsectionmark##1{}}

\oddsidemargin=\evensidemargin
\newcounter{sectionc}\newcounter{subsectionc}\newcounter{subsubsectionc}
\renewcommand{\section}[1] {\vspace{12pt}\addtocounter{sectionc}{1} 
\setcounter{subsectionc}{0}\setcounter{subsubsectionc}{0}\noindent 
        {\tenbf\thesectionc. #1}\par\vspace{5pt}}
\renewcommand{\subsection}[1] {\vspace{12pt}\addtocounter{subsectionc}{1} 
        \setcounter{subsubsectionc}{0}\noindent 
        {\bf\thesectionc.\thesubsectionc. {\kern1pt \bfit #1}}\par\vspace{5pt}}
\renewcommand{\subsubsection}[1] {\vspace{12pt}\addtocounter{subsubsectionc}{1}
        \noindent{\tenrm\thesectionc.\thesubsectionc.\thesubsubsectionc.
        {\kern1pt \tenit #1}}\par\vspace{5pt}}
\newcommand{\nonumsection}[1] {\vspace{12pt}\noindent{\tenbf #1}
        \par\vspace{5pt}}

\newcounter{appendixc}
\newcounter{subappendixc}[appendixc]
\newcounter{subsubappendixc}[subappendixc]
\renewcommand{\thesubappendixc}{\Alph{appendixc}.\arabic{subappendixc}}
\renewcommand{\thesubsubappendixc}
        {\Alph{appendixc}.\arabic{subappendixc}.\arabic{subsubappendixc}}

\renewcommand{\appendix}[1] {\vspace{12pt}
        \refstepcounter{appendixc}
        \setcounter{figure}{0}
        \setcounter{table}{0}
        \setcounter{lemma}{0}
        \setcounter{theorem}{0}
        \setcounter{corollary}{0}
        \setcounter{definition}{0}
        \setcounter{equation}{0}
        \renewcommand{\thefigure}{\Alph{appendixc}.\arabic{figure}}
        \renewcommand{\thetable}{\Alph{appendixc}.\arabic{table}}
        \renewcommand{\theappendixc}{\Alph{appendixc}}
        \renewcommand{\thelemma}{\Alph{appendixc}.\arabic{lemma}}
        \renewcommand{\thetheorem}{\Alph{appendixc}.\arabic{theorem}}
        \renewcommand{\thedefinition}{\Alph{appendixc}.\arabic{definition}}
        \renewcommand{\thecorollary}{\Alph{appendixc}.\arabic{corollary}}
        \renewcommand{\theequation}{\Alph{appendixc}.\arabic{equation}}
        \noindent{\tenbf Appendix \theappendixc #1}\par\vspace{5pt}}
\newcommand{\subappendix}[1] {\vspace{12pt}
        \refstepcounter{subappendixc}
        \noindent{\bf Appendix \thesubappendixc. {\kern1pt \bfit #1}}
        \par\vspace{5pt}}
\newcommand{\subsubappendix}[1] {\vspace{12pt}
        \refstepcounter{subsubappendixc}
        \noindent{\rm Appendix \thesubsubappendixc. {\kern1pt \tenit #1}}
        \par\vspace{5pt}}

\newcommand{\textlineskip}{\baselineskip=13pt}
\newcommand{\smalllineskip}{\baselineskip=10pt}

\def\eightcirc{
\begin{picture}(0,0)
\put(4.4,1.8){\circle{6.5}}
\end{picture}}
\def\eightcopyright{\eightcirc\kern2.7pt\hbox{\eightrm c}} 

\newcommand{\copyrightheading}[1]
        {\vspace*{-2.5cm}\smalllineskip{\flushleft
        {\footnotesize International Journal of Modern Physics C, #1}\\
        {\footnotesize $\eightcopyright$\,\,\, World Scientific Publishing
         Company}\\
         }}

\newcommand{\publisher}[2]{{\begin{center}\footnotesize\smalllineskip 
        Received #1\\
        Revised #2
        \end{center}
        }}

\def\abstracts#1#2#3{{
        \centering{\begin{minipage}{4.5in}\baselineskip=10pt\footnotesize
        \parindent=0pt #1\par
        \parindent=15pt #2\par
        \parindent=15pt #3\par
        \end{minipage}}\par}} 

\renewenvironment{thebibliography}[1]
        {\frenchspacing
         \ninerm\baselineskip=11pt
         \begin{list}{\arabic{enumi}.}
        {\usecounter{enumi}\setlength{\parsep}{0pt}     
         \setlength{\leftmargin 17pt}{\rightmargin 0pt}   
         \setlength{\itemsep}{0pt} \settowidth
        {\labelwidth}{#1.}\sloppy}}{\end{list}}

\newcounter{itemlistc}
\newcounter{romanlistc}
\newcounter{alphlistc}
\newcounter{arabiclistc}

\newcommand{\fcaption}[1]{
        \refstepcounter{figure}
        \setbox\@tempboxa = \hbox{\footnotesize Fig.~\thefigure. #1}
        \ifdim \wd\@tempboxa > 5in
           {\begin{center}
        \parbox{5in}{\footnotesize\smalllineskip Fig.~\thefigure. #1}
            \end{center}}
        \else
             {\begin{center}
             {\footnotesize Fig.~\thefigure. #1}
              \end{center}}
        \fi}

\newcommand{\tcaption}[1]{
        \refstepcounter{table}
        \setbox\@tempboxa = \hbox{\footnotesize Table~\thetable. #1}
        \ifdim \wd\@tempboxa > 5in
           {\begin{center}
         \parbox{5in}{\footnotesize\smalllineskip Table~\thetable. #1}
            \end{center}}
        \else
             {\begin{center}
             {\footnotesize Table~\thetable. #1}
              \end{center}}
        \fi}

\def\@citex[#1]#2{\if@filesw\immediate\write\@auxout
        {\string\citation{#2}}\fi
\def\@citea{}\@cite{\@for\@citeb:=#2\do
        {\@citea\def\@citea{,}\@ifundefined
        {b@\@citeb}{{\bf ?}\@warning
        {Citation `\@citeb' on page \thepage \space undefined}}
        {\csname b@\@citeb\endcsname}}}{#1}}

\newif\if@cghi
\def\cite{\@cghitrue\@ifnextchar [{\@tempswatrue
        \@citex}{\@tempswafalse\@citex[]}}
\def\citelow{\@cghifalse\@ifnextchar [{\@tempswatrue
        \@citex}{\@tempswafalse\@citex[]}}
\def\@cite#1#2{{$\null^{#1}$\if@tempswa\typeout
        {IJCGA warning: optional citation argument 
        ignored: `#2'} \fi}}

\def\pmb#1{\setbox0=\hbox{#1}
        \kern-.025em\copy0\kern-\wd0
        \kern.05em\copy0\kern-\wd0
        \kern-.025em\raise.0433em\box0}

\def\fnt#1#2{\footnotetext{\kern-.3em
        {$^{\mbox{\scriptsize #1}}$}{#2}}}

\def\fpage#1{\begingroup
\voffset=.3in
\thispagestyle{empty}\begin{table}[b]\centerline{\footnotesize #1}
        \end{table}\endgroup}

\def\runninghead#1#2{\pagestyle{myheadings}
\markboth{{\protect\footnotesize\it{\quad #1}}\hfill}
{\hfill{\protect\footnotesize\it{#2\quad}}}}
\font\tenbf=cmbx10
\font\tenit=cmti10 
\font\tenit=cmti10
\font\bfit=cmbxti10 at 10pt

\font\ninerm=cmr9

\font\eightrm=cmr8

\def\lsym{\raise-3pt\hbox{\vbox{\tabskip0pt\offinterlineskip
        \halign{\tabskip0pt plus 1em
        ##\tabskip0pt\cr
        $\,\,<\,\,$\cr
        $\,\,\sim\,\,$\cr}}}}
\def\rsym{\raise-3pt\hbox{\vbox{\tabskip0pt\offinterlineskip
     \halign{\tabskip0pt plus 1em
      ##\tabskip0pt\cr
      $\,\,>\,\,$\cr
      $\,\,\sim\,\,$\cr}}}}
\def\qed{\hbox{${\vcenter{\vbox{                        
        \hrule height 0.4pt\hbox{\vrule width 0.4pt height 6pt
        \kern5pt\vrule width 0.4pt}\hrule height 0.4pt}}}$}}
\def\theequation{\thesection.\arabic{equation}}         

\begin{document}
\parindent0em
\def\vm{v_{max}}
\def\q{\bar{p}}
\newcommand{\btau}{\boldsymbol{\tau}}
\newcommand{\bttau}{\boldsymbol{\tilde{\tau}}}
\newcommand{\btaujn}{\boldsymbol{\tau}_j^{(n)}}
\newcommand{\balpha}{\boldsymbol{\alpha}}
\newcommand{\bbeta}{\boldsymbol{\beta}}
\newcommand{\nn}{{\cal N}}
\newcommand{\Pt}{\tilde{\mathbf P}}
\newcommand{\cP}{{\cal P}}
\newcommand{\be}{\begin{equation}}
\newcommand{\ee}{\end{equation}}
\newcommand{\bea}{\begin{eqnarray}}
\newcommand{\eea}{\end{eqnarray}}
\newcommand{\nonu}{\nonumber\\}


\runninghead{G. Diedrich, L. Santen, A. Schadschneider \& J. Zittartz}
{Effects of on- and off-ramps}

\normalsize\textlineskip
\thispagestyle{empty}
\setcounter{page}{1}

\copyrightheading{Vol. 0, No. 0 (2000) 000--000}

\vspace*{0.88truein}

\fpage{1}
\centerline{\bf EFFECTS OF ON- AND OFF-RAMPS IN CELLULAR }
\vspace*{0.035truein}
\centerline{\bf AUTOMATA MODELS FOR TRAFFIC FLOW}
\vspace*{0.37truein}
\centerline{\footnotesize G. Diedrich, L. Santen, A. Schadschneider,
and J. Zittartz} 
\vspace*{0.015truein}
\centerline{\footnotesize\it Institut f\"ur Theoretische Physik,
  Universit\"at zu K\"oln,} 
\baselineskip=10pt
\centerline{\footnotesize\it D--50937 K\"oln, Germany}

\vspace*{0.225truein}
\publisher{\today}

\vspace*{0.21truein}
\abstracts{
We present results on the modeling of on- and off-ramps in 
cellular automata for traffic flow, especially the Nagel-Schreckenberg model. 
We study two different types of on-ramps that cause qualitatively the same 
effects. In a certain density regime $\rho_{low} <\rho <\rho_{high}$ one 
observes plateau formation in the fundamental diagram. The plateau 
value depends on the input-rate of cars at the on-ramp. The on-ramp acts as a 
local perturbation that separates the system into two regimes: A regime of free
flow and another one where only jammed states exist. This 
phase separation is the reason for the plateau formation and implies a 
behaviour analogous to that of stationary defects. This analogy allows to 
perform very fast simulations of complex traffic networks with a large number 
of on- and off-ramps because one can parametrise on-ramps in an exceedingly 
easy way.
}{}{}

\vspace*{10pt}


\vspace*{1pt}\textlineskip      
\section{Introduction}          
\label{intro}
\vspace*{-0.5pt}
An effective management of transport problems is of great interest in
densely populated areas. Therefore the  performance optimisation of the
existing traffic networks is an important aspect of infra-structural
planning \cite{nagel99}. Recent experimental
\cite{Daganzo99,Kerner961,Kerner972,Kerner981,Neubert99} 
and theoretical studies
\cite{leeetal,hellprl1,gunter}
have provided great evidence that the network
performance is large\-ly dominated by the capacity of so-called 
bottlenecks, i.e. parts of the roads where the capacity is locally
reduced. In realistic traffic systems a large variety of possible bottlenecks
exists, e.g. road constructions, crossings and lane-reductions. 
If one concentrates on highway networks the reduction of the capacity is
often due to on- and off-ramps. Therefore the influence of on- and 
off-ramps has been discussed, e.g., on- and off-ramps have been
used in studies of macroscopic models in order to explain the emergence of 
synchronised traffic \cite{leeetal,hellprl1}.

In this work we present a simulation study of on- and off-ramps using a
discrete microscopic traffic model \cite{NagelS,Ito} (for reviews, 
see \cite{juelich,tgf97,chowd}). In order to extract
the effects of  on- and off-ramps we keep the lattice geometry as simple as 
possible, i.e. we study a single-lane system with periodic boundary
conditions, with an additional on- and off-ramp. Moreover we
consider only one type of cars. This choice of the system allows 
for an easy parametrisation and systematic analysis of ramp effects.
Nevertheless these simplifications do not reduce the
practical relevance of our results because the empirical results
which have been obtained at multi-lane highways show the large
influence of the ramps \cite{Kerner961,Kerner972,Neubert99}. Moreover
the strong coupling of the lanes at 
high densities implies that the generic behaviour is found on all 
lanes \cite{Debch,Knospe,HH}.

The simulation results show that the maximal capacity of the system
depends on the flow at the ramps as well as on different input
strategies. In contrast, different output procedures do not affect the
results significantly. Starting from this realistic scenario we discuss the
relation to periodic systems with stationary defects \cite{lebo,defect},
 i.e. to periodic systems
where the mobility of the cars is locally reduced. It turns out
that the essential features of traffic systems near ramps are already
reproduced by the system with a stationary defect. Moreover we are 
able to establish a direct relation between on-ramp activity and
parametrisation of the defect. Therefore it seems to be possible to
achieve a considerable reduction of the computational complexity of
network simulation using this kind of implementation of ramps.

The influence of on- and off-ramps on the dynamics in microscopic
models is not yet analysed to our knowledge. The most important point
that is investigated here is the 
strategy of the cars to change from the acceleration lane to the driving lane.
This strategy depends on the individual behaviour of the drivers and might 
have a considerable influence on the traffic dynamics. In this paper we 
will present a simple scenario that is able to quantify the effects of on- and 
off-ramps qualitatively. It will be no restriction to the problem that we will 
only treat one-lane models. The effects
caused by the on-ramps in realistic systems can be explained 
qualitatively with one-lane models.
Therefore the Nagel-Schreckenberg (NaSch) model serves as the basis model
for the implementation of on- and off-ramps. Firstly we suggest two different 
types of on-ramps. It is found that it is not necessary to make a 
difference in the modelling 
of the off-ramps because they play a rather inferior role for the traffic 
dynamics. Afterwards we study the effect of both types on the model 
qualitatively with the help of suitable observables. It will be shown that 
there exists a strict analogy of the effect of on-ramps to that of stationary 
defects. This analogy allows the parametrisation of on-ramps in an exceedingly
easy way, so that one can perform very fast, but realistic, simulations of 
complex traffic networks.

Before we start with our considerations concerning the on- and off-ramps, let 
us remind the reader of the update rules (parallel dynamics) in the NaSch 
model:
\begin{description}
\item[{\bf R1}] Acceleration:\ \ \ \ \ \ \ $v_i \to \min(v_{max},v_i+1)$
\item[{\bf R2}] Braking:\ \ \ \ \ \ \ \ \ \ \ \ \ $v_i \to \min(d_i,v_i)$
\item[{\bf R3}] Randomization:\ \ \ \ $v_i \to \max(0,v_i-1)$\\ 
\phantom{Randomization:\ \ \ \ \ }with probability $p$
\item[{\bf R4}] Driving:\ \ \ \ \ \ \ \ \ \ \ \ \ \ car $i$ moves $v_i$ cells
\end{description}
Here $v_i$ and $d_i$ denotes the velocity of car $i$ and the number 
of empty cells in front of car $i$, i.e.\ the so-called headway, respectively. 
The maximum velocity and the slowdown parameter are denoted as $\vm$ and $p$ 
respectively.


\section{Definition of the ramp-types}
\label{ramp_types}

In this section we discuss two possible types of implementations of on- and 
off-ramps. The on- and off-ramps are implemented as connected parts of
the lattice where the vehicles may enter or leave the system. The
activity of the ramps is characterised by the number of entering (or
leaving) cars per unit of time  $j_{in}$ ($j_{out}$).  In order to
avoid density fluctuations we only added a car to the system if the
removal of another car at the off-ramp is possible at the {\em same}
time-step. So, the following relation for the rates holds: $j_{in}=j_{out}$.
Moreover, input and output are performed with a constant
frequency. Compared to a stochastic in- and output of cars this
particular choice allows for a better quantitative analysis of the 
results. We have tested that a stochastic in- and output of cars does 
not lead to a qualitatively different behaviour of the system.    
The chosen length of the on- and off-ramps and the distance between
them is motivated by the dimensions found on german highways. Here we
have chosen $L_{ramp}=25$ as length of the ramps in units of the lattice 
constant (usually identified with 7.5 meters). The 
position, i.e. the first cell of the on-ramp, is located at
$x_{on}=80$ and that of  
the off-ramp at $x_{off}=L-80$ where $L$ is the system size. Using  
periodic boundary conditions the distance of the on-ramp to the off-ramp is 
given by $d_{ramp}=x_{on}-x_{off}+L$. The only and essential difference in the 
implementation of the different types of on-ramps lies in the strategy of 
the cars changing from the acceleration lane to the driving lane.

Now we discribe two different procedures adding cars to the lattice.\\

{\bf Type A}
\vspace{1.5mm}

\begin{figure}[ht]
\centerline{\epsfig{figure=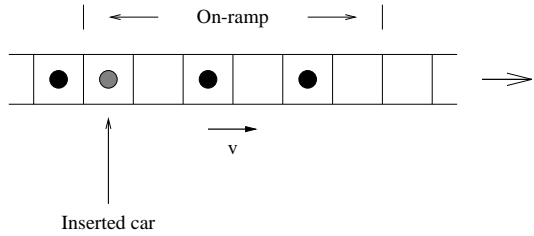,height=3cm}}
\caption{Schematical representation of the on-ramp (Type A).}
\label{on_ramp}
\end{figure}

Using this method the lattice will successively be searched in the region 
of the on-ramp ($x_{on}$ to $x_{on}+L_{ramp}$) until a vacant cell is found. 
Then a car will be inserted into this 
cell (see Fig.~\ref{on_ramp}), even if the cell in front is already occupied 
by a car. The velocity of the car is set to $v_{max}$. Dependent on the 
global density $\rho$ this can lead to a strong 
perturbation of the system (see Section \ref{effects}).

This extremely restrictive method described above implies a very
inconsiderate behaviour of the drivers because in the process of changing 
from the acceleration lane to the driving lane no safety margin will be held. 
However, due to the unbounded deceleration capabilities of the cars 
in the NaSch model this will not lead to accidents.

\pagebreak

{\bf Type B}
\vspace{1.5mm}

In this method the cell that will be occupied by a car is selected in a 
stochastic way. A vacant cell will be randomly chosen in the region of the 
on-ramp. Afterwards a car will be inserted into this cell like in 
type A. The measurements in the simulations will show that there is no 
qualitative difference of this type to type A. Type B will also lead to a 
strong perturbation of the system.

By the method described above one gets a more realistic description of traffic 
flow than in type A. Here it will be used to simulate both inconsiderate and 
cautious behaviour of the drivers.

\begin{figure}[ht]
\centerline{\epsfig{figure=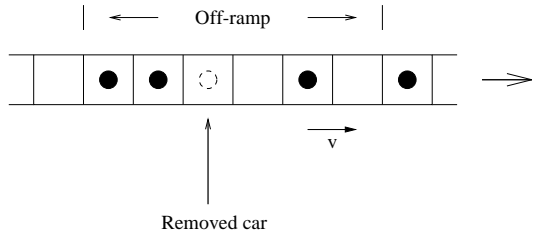,height=3cm}}
\caption{Schematical representation of the off-ramp.}
\label{off_ramp}
\end{figure}

The off-ramps work for the corresponding types of on-ramps in the same way: 
one goes successively through the cells of the lattice in the region of the 
off-ramps until an occupied cell is found. Then the car will be removed 
from this cell. A schematic diagram of this procedure is shown in 
Fig.\ref{off_ramp}.

\setcounter{section}{3}
\setcounter{equation}{0}
\section{Effects of the ramps}
\label{effects}

In this section we discuss the effects of the ramps. Using 
computer simulations fundamental diagrams dependent on the in/output-rates 
and density profiles dependent on the global density of the system are 
generated. Then, by analysing these results, we investigate the relation 
between on-ramps and stationary defects. This will be done in the last 
subsection.

\subsection{Fundamental diagrams and density profiles}
\label{fund_prof}

\begin{figure}[ht]
\centerline{\epsfig{figure=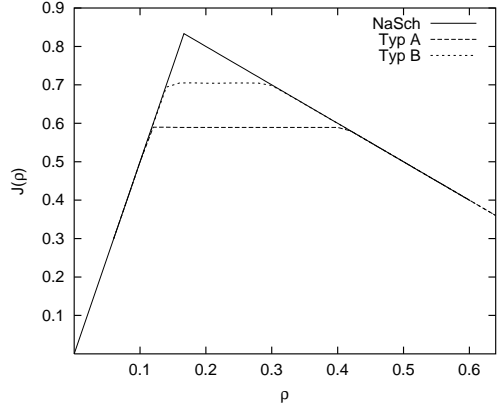,height=5.5cm}}
\caption{Fundamental diagrams of both ramp types at the same input-rate $j_{in}=\frac{1}{5}$. The model parameters are given by $L=3000, v_{max}=5, p=0$.}
\label{fd_ramp}
\end{figure}

In Fig.~\ref{fd_ramp} the fundamental diagrams of the NaSch model for both 
types of on-ramps are shown.
The input-rate is chosen to be $j_{in}=j_{out}=\frac{1}{5}$. This choice 
guarantees the conservation of the global density $\rho$. For comparison the 
fundamental diagram without ramps is also shown as solid line.

It is clearly seen that a density regime $\rho_{low}<\rho<\rho_{high}$ 
exists where the flow $J(\rho)$ is independent 
of the density. This so-called plateau value of the flow is lower than the
corresponding flow of the model without ramps. 
An increase of the input-rate $j_{in}$ leads to a decrease of the plateau 
value
If one compares the different types of input strategies it is evident
that the plateau value of type B is lower than that of  
type A. This can be explained by the probability that the cell in front of 
the inserted car is already occupied. For type B this probability is 
smaller than for type A. 
Nevertheless there is no difference between the two types in a qualitative
sense. That is the reason why we restrict ourselves to type A in the 
further treatment.
With this type the effect can be seen in the simplest way and so it is easier 
to describe.

Obviously we can distinguish three different phases depending on the global 
density. In the high and low density phases the 
average flow $J(\rho)$ of the perturbated system takes the same value as in 
the system without ramps. For intermediate densities 
$\rho_{low}<\rho<\rho_{high}$ the flow is constant and limited by the capacity 
of the ramps.

\begin{figure}[ht]
\centerline{\epsfig{figure=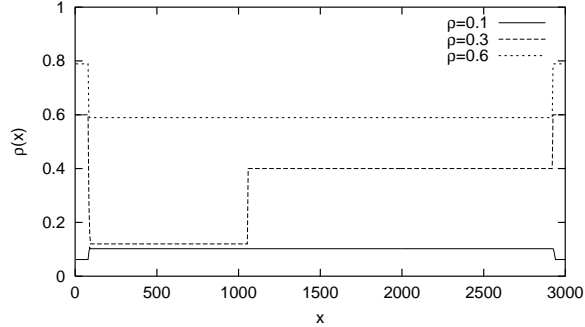,height=4.5cm}}
\caption{Density profiles in the three phases. In the high ($\rho=0.60$) and 
low ($\rho=0.10$) density phases only local inhomogeneities occur near the 
ramps ($x_{on}=80, x_{off}=L-80$). For intermediate densities ($\rho=0.30$) 
one observes phase separation. The model parameters are the same we used in 
Fig.~\ref{fd_ramp}.}
\label{dp_ramp}
\end{figure}

\begin{figure}[ht]
\vspace{-2mm}
\centerline{\epsfig{figure=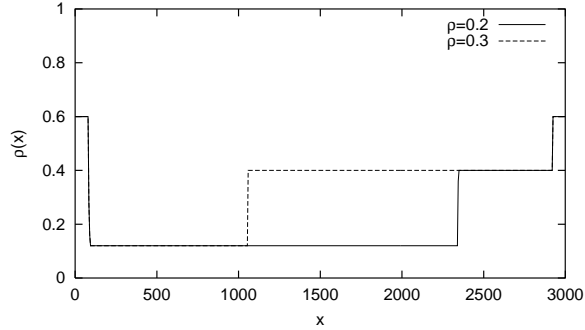,height=4.5cm}}
\caption{Density profiles for two different densities within the 
phase-separated state. $\rho_{ramp}, \rho_{low}, \rho_{high}$ remain constant,
only the length of the regions changes.}
\label{dp_ramp_2}
\end{figure}

For a better understanding of the behaviour of the average flow it will be 
helpful to look at the density profile (see Fig. \ref{dp_ramp}). In the high 
and low density phase one only observes local deviations from a constant 
density profile. For intermediate densities the system is separated into 
macroscopic high ($\rho_{high}$) and low ($\rho_{low}$) density regions. 
In the region of the ramps the density $\rho_{ramp}$ is additionally higher 
than in the high density region. Thus the ramps act like a blockage in the 
system that decreases the flow locally. Varying the global density within the 
phase-separated state the bulk densities in the high and low density region 
remain constant, only the length of the regions changes 
(see Fig. \ref{dp_ramp_2}). Also the local 
density at the ramps ($\rho_{ramp}$) remains constant. Finally this leads to 
a constant flow in the segregated phase.


If one assumes that the cars can move freely in the low density
regime and that the flow in the high density region depends
linearly on the density (which holds for small braking parameters
$p$) we get the following relation: 
\begin{equation}
\rho_{low}=\left(1-\rho_{high}\right)\frac{1-p}{v_{max}-p}.
\end{equation}
Unfortunately no expressions for 
$\rho_{low}$ and $\rho_{high}$ depending on the model parameters only can be 
given. The reason is that no exact analytical description for the 
unperturbated system with $v_{max}>1$ has been found so far. Furthermore an 
analytical treatment of the ramps is very complicated because of the strong 
interactions within these regions. 

\subsection{Analogy to stationary defects}
\label{analogy}

In the previous subsection we have shown that the on-ramps in the system act 
like a blockage. This blockage leads to a decrease of the local flow at the 
ramp. To clarify the analogy of on-ramps to stationary defects we give a short 
summary of the most important results on stationary defects like construction 
sites \cite{lebo,defect}.

\begin{figure}[ht]
\centerline{\epsfig{figure=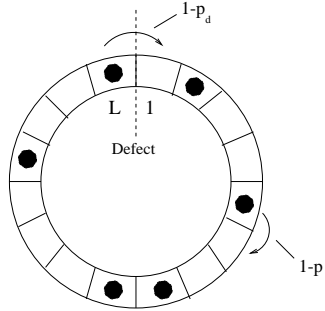,height=4.1cm}}
\caption{Schematical representation of a single defect in the NaSch model 
with $v_{max}=1$. In this special case the average velocity of the cars is 
approximately $v=1-p$ for $\rho\ll\frac{1}{2}$. At the defect site the 
average velocity is $v=1-p_{d}$.}
\label{defect}
\end{figure}

Stationary defects can be implemented in two different ways. First, one can 
define a certain range on the lattice where the slowdown parameter $p$ is 
increased compared to that of the residual lattice sites \cite{defect}. 
Second, one can decrease the maximum velocity $v_{max}$ in this certain range 
\cite{emmer,vicsek}. 
In our investigations we restrict ourselves to the first method.

Fig.~\ref{defect} shows a schematical representation of a single defect in the 
NaSch model with $v_{max}=1$ and slowdown parameter $p$. The slowdown parameter
at the defect is denoted as $p_d$.

\begin{figure}[ht]
\centerline{\epsfig{figure=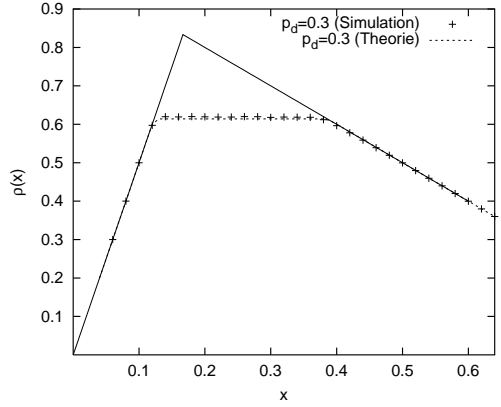,height=5.5cm}}
\caption{Fundamental diagram of the NaSch model with stationary defects. The 
model parameters are given by $L=3000, p=0, v_{max}=5$. For comparison the 
diagram of the model without defects is shown as solid line.}
\label{fd_def}
\end{figure}

In Fig. \ref{fd_def} the fundamental diagram of the NaSch model with 
stationary defects is represented. Again we can distinguish three different 
phases. In the high and low density regime the flow $J(\rho)$ is identical to 
that of the model without defects. For intermediate densities 
$\rho_{low}<\rho<\rho_{high}$ the flow is constant in analogy to the model 
with ramps. In this regime $J(\rho)$ is limited by the capacity of the defect 
sites.

\begin{figure}[ht]
\centerline{\epsfig{figure=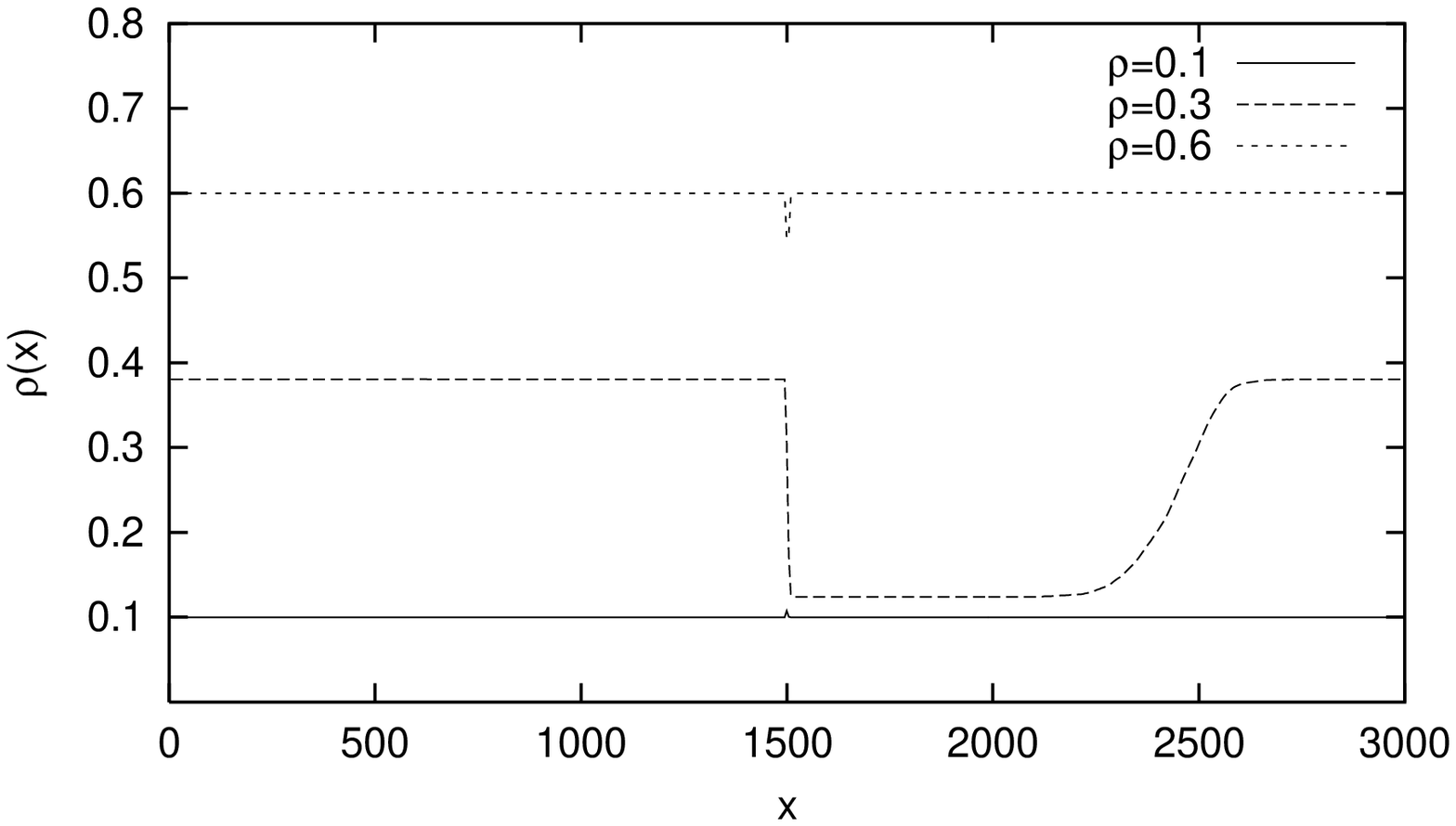,height=4.5cm}}
\caption{Density profiles of the NaSch model with stationary defects for 
densities in the three phases. In the high ($\rho=0.60$) and low ($\rho=0.10$)
density phases only local inhomogeneities occur near the defect sites 
($L-v_{max}\le x_d<L$). For densities within the plateau regime one observes 
phase separation.}
\label{dp_def}
\end{figure}

Once again the phenomenon of the plateau formation can be understood by the 
analysis of the density profiles (see Fig.~\ref{dp_def}). The argument is 
the same as in subsection 3.1.

For the case $v_{max}=1$ of the NaSch model with a single defect site good 
approximations for the plateau value of the flow as well as for the densities 
$\rho_{low}$ and $\rho_{high}$ can be made \cite{defect}. An essential reason 
for this is the particle-hole symmetry in this special case. Because of this 
symmetry following relation holds:
\begin{equation}
\label{p_h_sym}
\rho_{low}+\rho_{high}=1.
\end{equation}
With the assumption \cite{lebo} that the system is separated into two regions 
of constant density and that the flow $J_d$ at the defect is equal to the flow 
$J_{bulk}$ in the bulk one can establish following equation in the 
mean field theory:
\begin{equation}
J_d=q_d\rho_{high}\left(1-\rho_{low}\right)=J_{bulk}=q\rho_{high}
\left(1-\rho_{high}\right) \nonumber
\end{equation}
where $q=1-p$ and $q_d=1-p_d$. With equation (\ref{p_h_sym}) we finally get 
the expressions for the densities:
\begin{equation}
\rho_{high}=\frac{q}{q+q_d}\quad,\quad\rho_{low}=\frac{q_d}{q+q_d}.
\end{equation}
From this and the exact current (see e.g. \cite{Ito,shad99}) follows 
the plateau value of the flow:
\begin{equation}
J_P=\frac{1}{2}\left(1-\frac{1}{q+q_d}\sqrt{{\left(q+q_d\right)}^2-4q^2q_d}
\right).
\end{equation}
In principle one can treat the defects for $v_{max}>1$ in the same way. When we
make the assumption that for the two branches in the fundamental diagram the
following approximations hold:
\begin{equation} 
J_{free}=\rho\left(v_{max}-p\right)\quad,\quad
J_{jam}=\left(1-p\right)\left(1-\rho\right), \nonumber
\end{equation}
one gets the following expressions for the densities in the phase-separated 
state:
\begin{eqnarray}
\rho_{low}  &=& \frac{q_d}{v_{max}-p+q_d},\nonumber\\
\rho_{high} &=& \frac{(v_{max}-p)(q-q_d)+qq_d}{(v_{max}-p)q+qq_d}.
\end{eqnarray}
The plateau value of the flow results from this as:
\begin{equation}
\label{flow_plat}
J_P=\frac{(v_{max}-p)q_d}{v_{max}-p+q_d}.
\end{equation}
In spite of the crude estimate for $J_{jam}$ this approximation is in good 
agreement with the simulation data for small $p$ (see Fig. \ref{fd_def}). It 
has not been checked so far for which range of values for $p_d$ equation 
(\ref{flow_plat}) can reproduce the simulation results.

We have shown that there is no qualitative difference between the effect of 
on- and off-ramps and that of stationary defects. In both cases one observes 
plateau formation in the fundamental diagram as well as phase separation in 
the system. The only difference lies in the nature of the blockage dividing
the system into two macroscopic regions. In the case of the ramps it is the 
local increase of the density that decreases the flow locally. In the model
with defects the increased slowdown parameter leads to a local decrease 
of the flow.

\begin{figure}[ht]
\centerline{\epsfig{figure=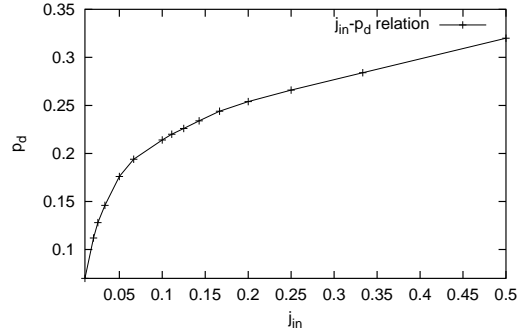,height=4.5cm}}
\caption{Relation between the input-rates $j_{in}$ and the slowdown 
parameter $p_d$ at the defect sites.}
\label{in_pd_dep}
\end{figure}

In order to use the equivalence of ramps and defects in realistic simulations 
of traffic flow one has to determine the relation between the input-rates 
$j_{in}$ and the slowdown parameter $p_d$ at the defect sites. 
Fig. \ref{in_pd_dep} shows the values of $p_d$ and $j_{in}$ leading to the 
same fundamental diagrams. It can be seen that a non-trivial relation between 
the two parameters exists. We regard this Fig. \ref{in_pd_dep} as our main 
result for future applications.



\section{Summary and Conclusions}
\label{sec_summary}
In this paper we have presented a simple scenario that can quantify
the effects  
of ramps on the traffic dynamics in microscopic models. The model used here is 
the well known Nagel-Schreckenberg model, a cellular automaton for traffic 
flow. We have introduced two different types of on-ramps. One type that 
implies a very inconsiderate behaviour of the drivers and another type that
simulates both inconsiderate and cautious behaviour. We have shown that 
there is no difference between these two types in a qualitative sense. With 
both types one observes effects very similar to those of stationary effects:
in both cases one observes plateau formation in the fundamental diagram as 
well as phase separation in the system. The only difference lies in the 
mechanism dividing the system into two macroscopic regions. This analogy is 
is very important for efficient modelling of complex networks 
\cite{nagel99} with a multitude of ramps.

Usually in simulations of large networks one uses a more realistic approach. 
In \cite{2lane} ramps are fed by an external road. Cars try to change to the 
highway according to the usual lane changing rules. If they could not change 
the cars wait at the end of the ramp.

In our considerations we have also investigated in the influence of ramps on 
different mutations of the NaSch model like the {\it Slow To Start} model 
{\cite{sts,barlovic} and a model with an anticipation rule. 
The effects caused by the ramps are qualitatively the same as in the 
NaSch model, but in the model with anticipation one has to use much higher 
input-rates $j_{in}$ to observe phase separation and plateau formation. 
This model reacts more robust on the perturbations caused by the cars 
changing from the acceleration lane to the driving lane \cite{diedrich}.

The effects of ramps have been studied in connection with synchronised traffic 
recently \cite{leeetal,hellprl1}. For the hydrodynamic models it has
been found  
that the localised perturbation through the ramp flow can induce transitions 
between different traffic states.

In \cite{gunter} it has been shown using computer simulations and data from 
measurements on real traffic that on-ramps can induce a first-order 
non-equilibrium phase transition between free flow and a congested phase. 
The transition is induced by the interplay of density waves caused by the 
on-ramps and shock waves moving on the highway.


\nonumsection{Acknowledgment}
\noindent
This work has been performed within the research program of the SFB 341 
(K\"oln--Aachen--J\"ulich).  L.~S. acknowledges support
from the Deutsche Forschungsgemeinschaft under Grant No. SA864/1-1.

%
%
\nonumsection{References}

\end{document}